\documentclass[preprint,proceedings]{rmaa}

\SetYear{2002} \SetConfTitle{Galaxy Evolution: Theory and
Observations}
\title{Modelling the Radio to X-ray SED of Galaxies}

\author{L. Silva\altaffilmark{1}, G.L. Granato\altaffilmark{2},  A. Bressan\altaffilmark{2},
  and P. Panuzzo\altaffilmark{3} }

\altaffiltext{1}{INAF - Osservatorio Astronomico di Trieste, Via
G.B. Tiepolo 11, I34131, Trieste, Italy (silva@ts.astro.it)}
\altaffiltext{2}{INAF - Osservatorio Astronomico di Padova, Italy}
\altaffiltext{3}{SISSA, Trieste, Italy}

\suppressfulladdresses

\listofauthors{L. Silva, G.L. Granato, A. Bressan, P. Panuzzo}
\indexauthor{Silva, L., Granato, G.L., Bressan, A., Panuzzo, P.}

\addkeyword{Infrared: galaxies} \addkeyword{Radio continuum: galaxies}
\addkeyword{X-rays: galaxies} 

\begin{document}
\maketitle

\boldabstract{Our multi-wavelength model GRASIL for the SED of galaxies is described, in particular the recent
extension to the radio and X-ray range.
With our model we can study different aspects of galaxy evolution
by exploiting all available spectral observations, 
where different emission components dominate.}

{\it\bf UV TO SUBMM:} The UV to submm model
is well established and has been applied in several
aspects of the study of galaxy evolution (Silva et al.\
1998, Bressan et al.\ 1998, Granato et al.\ 2000, 2001).

\begin{figure}[!t]
  \includegraphics[width=\columnwidth,height=7cm]{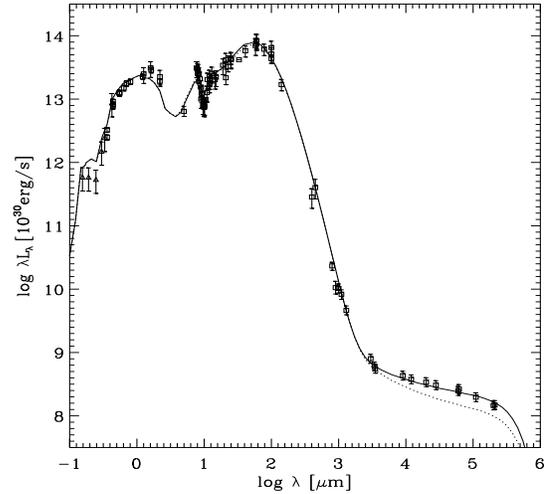}
  \vspace{-0.5truecm}
  \caption{Fit to the SED of M82: models with different starburst age are 
  degenerate in the UV to submm but not in the radio.}
  \label{figm82}
\end{figure}

{\it\bf RADIO:} We have extended our model to the radio range (Bressan, Silva \& Granato 2002) 
with both the thermal ($\propto$ ionizing luminosity) 
and the non thermal emission ($\propto$ SNRate). 
The radio emission coupled with the FIR provides a diagnostic tool
to constrain the age of starbursts (Figure~\ref{figm82}). 

\begin{figure}[!t]                                                                         
  \includegraphics[width=\columnwidth,height=8.7cm]{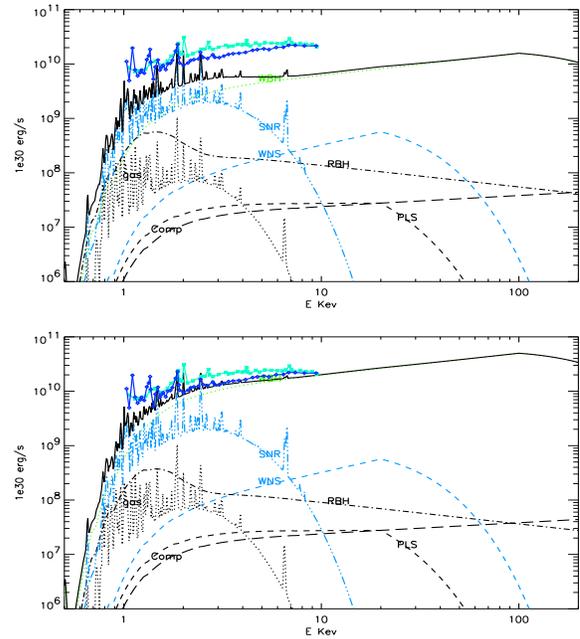}                                            
  \vspace{-0.5truecm}
  \caption{Above: minimum mass for BH formation=
  14 M$_\odot$, M$_{BH}=$ 5 M$_\odot$. WBH, WNS= wind-fed BH and NS, RBH=Roche-lobe fed BH, PLS=Pulsars, 
  SNR= supernova remnants. Gas and IC emissions are arbitrarily scaled. 
  Below: same but M$_{BH}=$ 10 M$_\odot$}                                                                
  \label{figxm82}                                                                            
\end{figure}

{\it\bf X-RAY:} We have included the contribution of stellar
populations to the X-ray emission of staburst galaxies. Following Van Bever \& Vanbeveren
(2000) we consider: Binaries with a NS or a BH as a primary, and a OB star as a secondary, Pulsars, SN Remnants.
The main uncertainties are the binary fraction, the minimum mass for BH formation, the initial
spin period of pulsars (whose $L_X \propto P_0^{-2}$), and the spectral shape to assign to the
different components. In Figure~\ref{figxm82} the X-ray SED expected for M82 
according to the model fitting its UV to radio SED is compared to observations
(Moran \& Lehnert 1997, Cappi et al. 1999). 

\end{document}